# On-chip plasmonic slit-cavity platform for room-temperature strong coupling with deterministically positioned colloidal quantum dots

Jin Qin,*†‡ Benedikt Schurr,†‡ Patrick Pertsch,† Daniel Friedrich,† Max Knopf,† Saeid Asgarnezhad-Zorgabad,¶ Lars Meschede, ¶ Daniel D.A. Clarke,¶ Monika Emmerling,† Artur Podhorodecki,§ Ortwin Hess, *¶ and Bert Hecht*†

†Nano-Optics and Biophotonics Group, Experimentelle Physik 5, Physikalisches Institut, Universität Würzburg & and Röntgen Research Center for Complex Material Research, Physics Institute, Am Hubland, Würzburg, D-97074, Germany.

¶School of Physics and CRANN Institute, Trinity College Dublin, Dublin, Ireland.

§Department of Experimental Physics, Wroclaw University of Science and Technology, Wybrzeze Wyspianskiego, Wroclaw, 50-370, Poland.

‡ These two authors contribute equally.
* jin.qin@uni-wuerzburg.de; Ortwin.Hess@tcd.ie; hecht@physik.uni-wuerzburg.de

ABSTRACT. Strong coupling between quantum emitters and optical cavities is essential for quantum information processing, high-purity single-photon sources, and nonlinear quantum devices. Achieving this regime at room temperature in a compact, deterministic on-chip platform—critical for integration with nanoelectronic circuitry and scalable device architectures—remains a major challenge, mainly due to the difficulty of fabricating cavities with ultra-small mode volumes and precisely positioning quantum emitters. Here, we demonstrate a robust quantum plasmonic device in which colloidal quantum dots (Qdots) are strongly coupled to plasmonic slit cavities using a dielectrophoresis-based positioning technique with real-time photoluminescence (PL) feedback, providing directly resolvable coupled structures that enable parallel device fabrication and straightforward integration with additional optical elements such as waveguides. Our measurements reveal clear PL-resolved Rabi splitting at room temperature with pre-characterized cavities, with variations across devices that scale with the average number of coupled Qdots. While electrical tuning via the quantum-confined Stark effect is enabled by integrated electrodes, its impact is largely overshadowed by room-temperature spectral diffusion. Our results pave the way for scalable, electrically tunable quantum plasmonic platforms, offering new opportunities for integrated quantum photonic circuits, active light-matter interactions, and room-temperature quantum technologies.

**Introduction**

The spontaneous emission of a quantum emitter can be significantly modified by the local density of optical states when coupled to a resonant optical mode, leading to rich light-matter interaction phenomena, such as Purcell-enhanced photon emission[1–6]. When emitter and cavity are in the strong-coupling regime, a new hybrid light-matter state, known as a polariton, emerges [7]. This state exhibits both photonic and excitonic characteristics and has enabled advances in areas including Bose-Einstein condensation[8], enhanced chemical reactivity[9], and potentially modified superconductivity[10].

Spectroscopically, strong coupling manifests as Rabi splitting in spectra of scattering, absorption, or photoluminescence (PL), indicating at least one complete cycle of coherent energy exchange between the emitter and the cavity. Achieving this regime requires a coupling strength $g$ that exceeds both the cavity loss and emitter decay rates including dephasing, which is especially challenging at room temperature. Dielectric cavities, with their high quality factors $Q$, are advantageous in terms of low loss[11,12]. However, to maximize the spectral overlap between such high-Q cavity resonances and quantum emitters, a cryogenic environment is typically required, as the emitter linewidth broadens with increasing temperature. The linewidths of the two coupled entities should be comparable to ensure sufficient spectral overlap, thereby maintaining the strong coupling condition.

In contrast, plasmonic cavities, consisting of nanoscale metallic structures, offer deep subwavelength confinement of optical modes with moderate $Q$ factors, yielding ultrafast energy transfer. This makes them promising candidates for achieving strong coupling at room temperature. However, realizing strong coupling in such systems demands precise spatial and spectral matching between the quantum emitter and the plasmonic cavity. In particular, optimal spatial matching poses a major challenge due to the need for nanometer-scale positioning of the emitter within the localized plasmonic field. Recent advancements have demonstrated various platforms and techniques to achieve strong coupling between quantum emitters and plasmonic cavities at room temperature, including particle-on-mirror structures[13–16], plasmonic tips[17,18], scanning slit cavities[19,20], bowtie antennas[21–23] and nanoparticles[24–27]. Methods such as scanning probes[17,18], or our own approach of integrating a slit cavity into an AFM tip[19,20], offer additional flexibility by leveraging the nanometer-scale precision of AFM scanners. Yet, such devices are inherently serial and not easily scalable. Particle-on-mirror structures, on the other hand, provide extremely small mode volumes through self-assembled junctions[13,15]. However, most implementations rely on drop-casting techniques[13–16,24–26], where quantum emitters are randomly deposited onto the plasmonic structures, resulting in non-deterministic placement and challenges in characterization. Moreover, while multi-stack architectures have been proposed for building devices[15,18], a planarized design is more favorable for integration with nanocircuitry or ultra-compact nanophotonic platforms—particularly when additional elements can be incorporated into the strongly coupled system to enable extended functionality, such as integration with waveguiding and gating structures. Despite the progress of these approaches, fully on-chip integration remains highly desirable because it combines robustness, scalability, and compatibility with established top-down nanofabrication methods, offering a clear pathway toward ultracompact quantum photonic devices.

Here, we develop an on-chip plasmonic platform consisting of a slit cavity structure and a counter electrode, as illustrated in Figure 1a. This open architecture allows retrofitting with a variety of quantum emitters and, in principle, can be integrated with additional waveguides. In this sense, it serves as a prototypical structure for devices whose functionality is based on cavity quantum electrodynamic effects, particularly strong coupling.. Using a dielectrophoresis (DEP) process, one or several quantum emitters can be positioned near the tip of the slit, where the localized plasmonic cavity mode exhibits a pronounced field maximum. Split PL spectra can be observed, indicating strong coupling. We find that the Rabi splitting strength varies between devices, which we attribute to differences in the specific quantum emitters coupled to the cavity mode. Additionally, we attempt to tune the emitter resonance via the quantum-confined Stark effect (QCSE) by applying an external voltage. However, due to significant spectral diffusion of the emitters at room temperature, the tunability is barely observable, with a typical decrease in PL intensity. Our experiments demonstrate on-demand strong coupling at room temperature in an ultracompact on-chip plasmonic device, highlighting its strong potential for integration in quantum photonic circuits.

**Results and Discussion**

**Plasmonic slit cavity characterization**

On-chip plasmonic slit cavity structures are fabricated using monocrystalline gold flakes with a thickness of 40 nm, employing helium ion milling (see scanning electron microscope (SEM) image and inset in Figure 1b). The nanoscale resolution and robustness of this fabrication method enable on-demand creation of individual well-defined structures with high reproducibility. The slit cavity supports Fabry-Pérot (FP) modes, which depend on the slit's length and width. As demonstrated in Figure 1c, using quasi-normal mode analysis (Supplementary Information S1), the resonance of second- and third-order FP modes characterized by the number of field maxima in the slit (Figure 1d), can be tuned by varying the slit length while keeping the width fixed. Besides, we also find that the free spectral range of the FP modes is sufficiently large to allow only a single FP mode to couple with the quantum emitter. Furthermore, both the second- and third-order FP modes exhibit quadrupolar-like field distributions (Figure 1d), which helps suppressing radiation losses and therefore yield relatively high $Q$.

In our experiment, we choose a slit length of 215 nm and a width of 7 nm to match the resonance of the quantum emitter. The slit width is comparable to the size of the emitter, enabling effective spatial overlap of the mode emanating at the tip and the quantum emitter. An additional counter electrode is fabricated opposing the tip of the slit cavity, with a gap of approximately 20 nm. This adds great versatility to the platform. For example, AC and DC voltages can be applied across the gap allowing for dielectrophoresis and to induce Stark shifts in emitters. The resulting FP mode in the plasmonic slit cavity is verified through photoluminescence (PL) spectroscopy under high excitation intensity, allowing both spatial and spectral characterization.

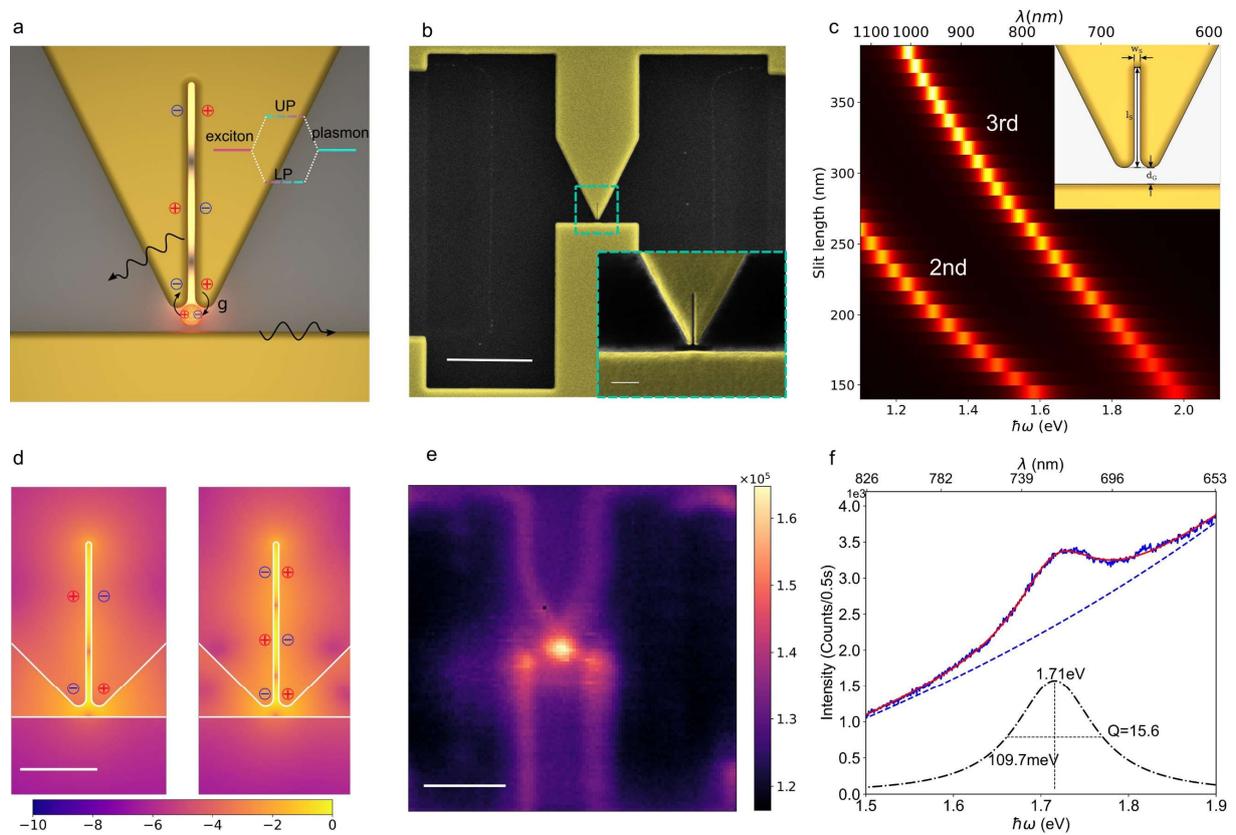

**Figure 1. Schematic of the on-chip plasmonic strong coupling platform and characterization of the plasmonic nanoslit cavity. a**, Illustration of a single quantum emitter interacting with an on-chip plasmonic nanoslit cavity. The third-order nanoslit cavity mode is strongly coupled with the quantum emitter. In the experiment, the emitter is first excited nonresonantly using a green laser. Due to coherent energy exchange, polaritons are formed that decay via emission of a photon.

UP: upper polariton and LP: lower polariton. **b**, SEM images of the on-chip strong coupling device. Scale bar: 1 μm. The inset shows a zoom to the tip area indicated by a dashed rectangle. Scale bar: 100 nm. **c**, Second- and third-order FP modes obtained through quasi-normal mode analysis (Supplementary Information S1) for varying slit lengths and a fixed slit width of 7 nm. **d**, Electric field distributions (log scale, $|E|^2$) of the second- and third-order FP modes in the nanoslit cavity. Scale bar: 100 nm. **e**, Hyperspectral PL map recorded by scanning the excitation spot across a sample region centered on the nanoslit cavity. Notably, this PL signal is only detectable under high laser power (wavelength: 532 nm, power: 500 μW, intensity: $4.4 \times 10^9 \text{W/m}^2$. The PL map is generated by integrating spectral intensities between 1.68 eV and 2.00 eV corresponding to the spectrum in **f**. Scale bar: 1 μm. **f**, PL spectrum (blue line) of the cavity mode when the excitation laser is focused on the slit structure. The spectrum is fitted with a cumulative model (red line) consisting of a Lorentzian resonance (black dot-dashed line, corresponding to the slit cavity mode) and an exponential background (blue dashed line). From the Lorentzian fit, the cavity resonance energy and quality factor are extracted.

To record spatially resolved PL spectra of the device shown in Figure 1b, we use hyperspectral imaging methodology with the experimental setup depicted in Figure S2 (Supplementary Information). Figure 1e shows a hyperspectral map collected by scanning the excitation spot across the entire device area. A bright localized spot appears at the plasmonic slit cavity position when the signal is integrated over the spectral range of its resonance. The corresponding PL spectrum recorded at the brightest spot is shown in Figure 1f, where a shoulder on the broad gold PL background indicates the slit cavity mode resonance. By applying a cumulative fit, we extract a cavity resonance at 1.71 eV with a quality factor Q = 15.6 (Supplementary Information S2).

**Quantum emitter characterization**

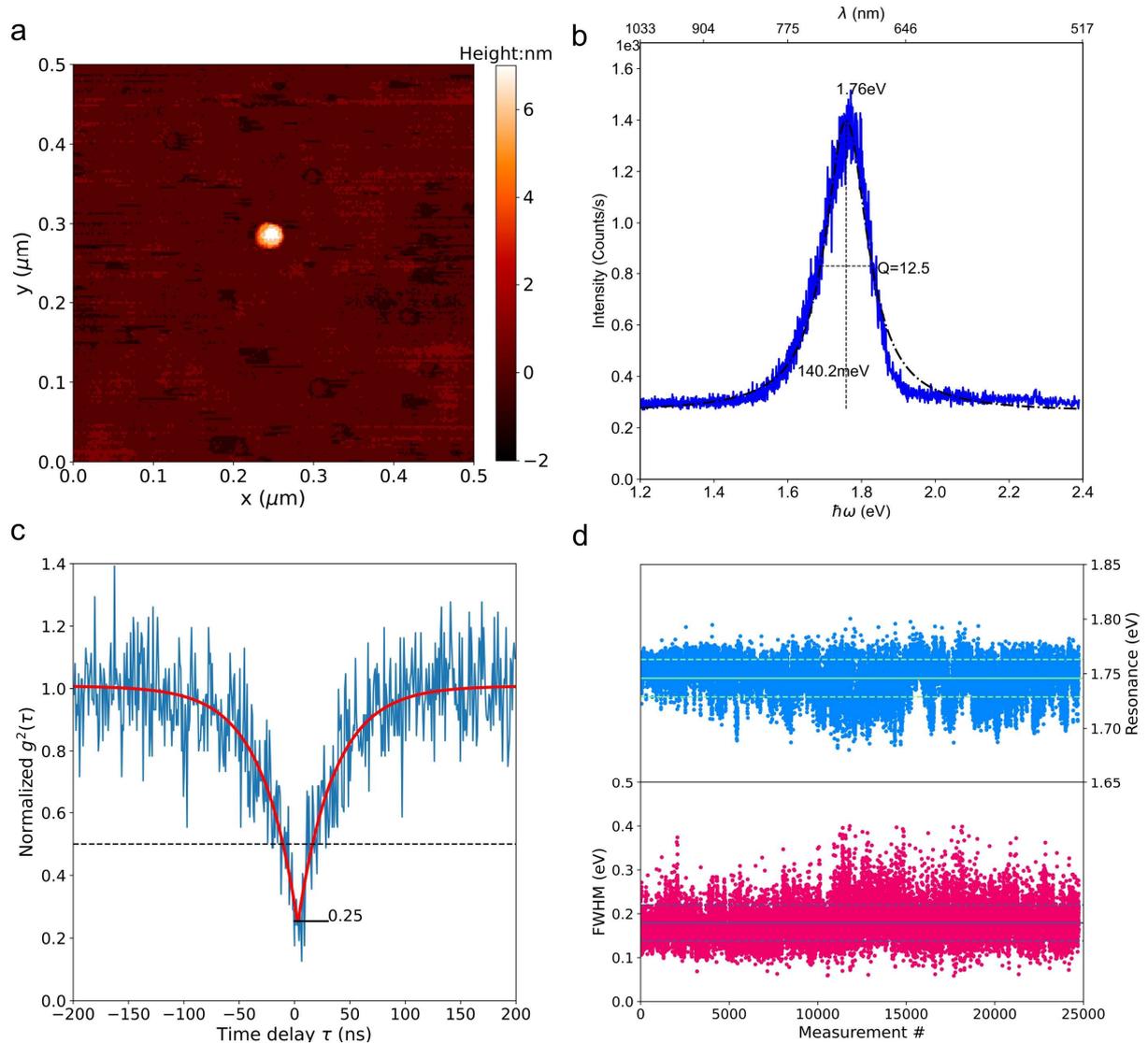

**Figure 2. Characterization of single Qdots. a**, AFM image of a representative single Qdot, spin-coated on a glass coverslip. **b**, PL emission spectrum (blue line) under green laser excitation with a power of 1 μW (intensity: $8.8 \times 10^6 W/m^2$). The spectrum is fitted with a Lorentzian function (black dot-dashed line) to extract the resonance energy and linewidth. **c**, Photon statistics of the Qdot emission measured by second-order autocorrelation. A clear antibunching dip of 0.25 at zero time delay confirms single-photon emission behavior. **d**, Time-resolved PL spectra of a single Qdot (Figure S3), recorded with a temporal resolution of 33 ms. Resonance energies (blue dots)

and linewidths (red dots) are extracted from Lorentzian fitting. The average values and one standard deviation are indicated by solid and dashed lines, respectively.

The quantum emitters used in our experiments are commercial colloidal core-shell quantum dots (Qdots, 705 nm, CdSeTe/ZnS, Thermo Fisher). These Qdots are first spin-coated onto a clean coverslip to characterize their physical and optical properties. The atomic force microscope (AFM) image in Figure 2a shows a typical single Qdot exhibiting an apparent diameter of approximately 15 nm and a height of approximately 7 nm, which closely matches the width of the slit cavity. The Qdot's emission spectrum, excited by a green laser, is displayed in Figure 2b. It can be fitted with a Lorentzian profile, yielding a resonance energy of 1.76 eV and a relatively broad emission linewidth of 140.2 meV. Single-photon emission by individual Qdots is confirmed by intensity auto-correlation measurements, as depicted in Figure 2c. Clear antibunching behavior is observed with a dip to 0.25 at zero time delay.

Under ambient conditions, some Qdots are chemically unstable and prone to oxidation when exposed to laser excitation resulting e.g. in a resonance shift[20]. To verify the absence of such behavior, we record time-traced PL spectra and extract the resonance energies and linewidths using Lorentzian fitting, as shown in Figure 2d and Figure S3. The data show no significant spectral shifts (i.e., no consistent blue or red shift), but we do observe substantial spectral diffusion at room temperature, with variations as large as 50 meV.

**DEP process**

To achieve strong coupling, one of the main challenges is positioning the quantum emitter with nanometer-scale precision to ensure spatial alignment with the localized cavity mode. To overcome this problem, we employ a DEP process with a real-time feedback mechanism to attract and position single or multiple Qdots at the tip of the slit cavity, where the cavity mode's plasmon-mediated electric field is also concentrated[28]. Prior to the DEP process, the Qdot solution is diluted in pure water at an appropriate ratio (1:10000). A 5 µL droplet of this diluted solution is then deposited onto the surface of the on-chip device. An alternating voltage is applied to the electrode containing the slit cavity, while the counter electrode is grounded (Figure 3a). This generates a non-uniform electric field that polarizes the quantum dots (Qdots) in solution, producing a net force that can either attract or repel the Qdots toward or away from the slit tip, depending on the applied frequency (Supplementary Information S4). By carefully tuning the applied voltage and frequency, the net force can be made attractive, drawing nearby Qdots toward the slit tip, where the electric field gradient is strongest.

To ensure successful placement of Qdots within the region of interest, we implement a real-time feedback mechanism based on PL detection using a high-bandwidth avalanche photodiode (APD). Before initiating the DEP process, we scan the excitation laser (532 nm, 1 µW) over the sample and record the resulting PL map (Figure 3c). From this map, the approximate position of the slit cavity can be identified, as indicated by the white dashed line. During DEP, the region of interest (slit cavity) is continuously monitored at a repetition rate of 100 Hz. As soon as Qdots are trapped at the cavity tip, a sharp increase in PL intensity is observed at that location (Figure 3d), signaling successful placement of Qdots. At this point, the DEP process is interrupted, and the device is allowed to dry for several minutes. The Qdots remain anchored at the desired location, as shown in Figure 3b. More details about the DEP process, its optimization and parameters used can be found in Supplementary Information S4.

The resulting coupled structures are first characterized using AFM. By varying the DEP parameters, such as applied frequency and voltages, we can achieve different numbers of captured

Qdots. In Figure 3e, it appears that only a single Qdot is attracted; however, this observation is not conclusive, as multiple Qdots could be stacked vertically, which is not resolvable by AFM. Additionally, the DEP process not only attracts Qdots but sometimes also contaminants, as observed in Figure 3f.

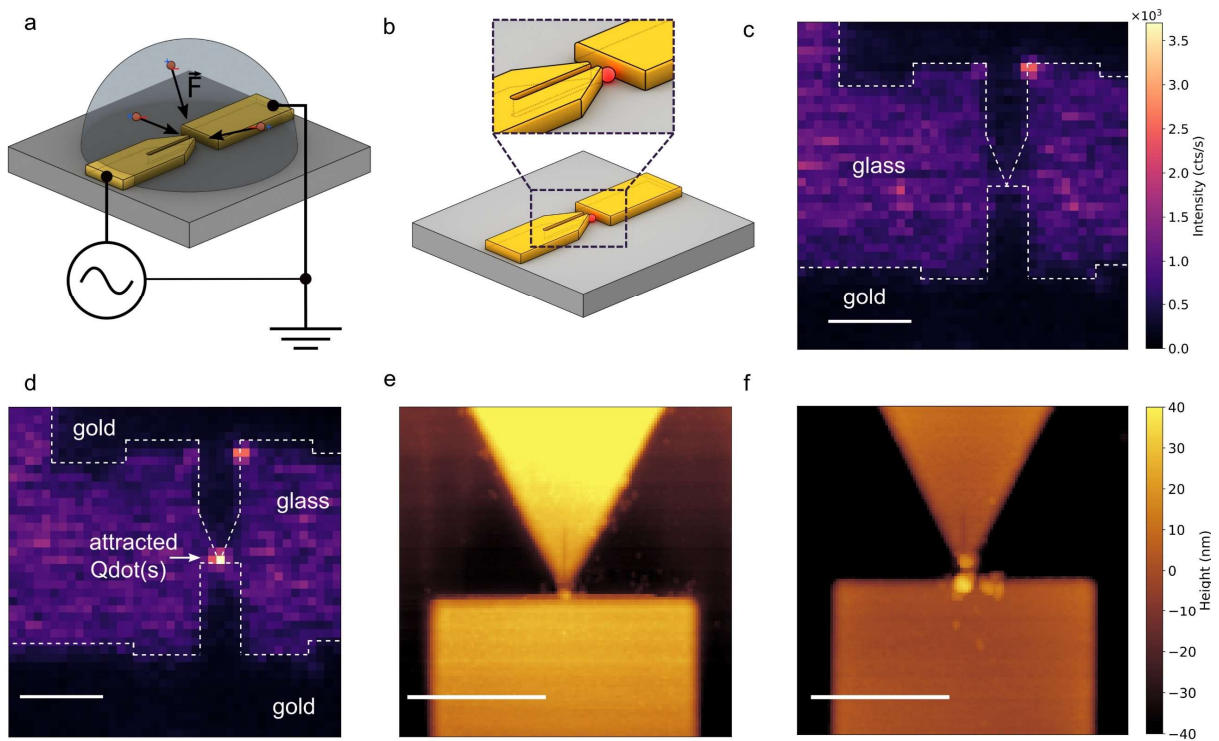

**Figure 3. DEP Process. a**, Schematic illustration of the setup used for DEP. A droplet of Qdot solution (diluted 1:10000 with pure water) is placed on top of the device, and an AC voltage (amplitude: 2V, frequency: 2 MHz) is applied across the plasmonic nanoslit cavity (left) and the grounded counter electrode (right). Under the influence of the non-uniform electric field, for the right range of frequencies, Qdots experience a dielectrophoretic force and migrate toward the tip of the slit cavity, where the field gradient is maximal. **b**, Illustration of the resulting coupled structure after DEP. A Qdot (represented by the red sphere) is positioned at the end of the slit

cavity. Inset: A zoom-in image. **c-d**, Real-time feedback mechanism based on PL signals detected by an APD, while scanning the excitation laser (power: 1 μW, intensity: $8.8 \times 10^6 \text{W/m}^2$) over the device. **c**, Prior to DEP, no confined PL spot is observed. Dark regions correspond to the gold surface, and the brighter areas correspond to weak autofluorescence of the underlying cover glass. Also, at these positions, more excitation light is transmitted and possibly also excites Qdots suspended in the solution, leading to background signals. **d**, Once one or more Qdots are captured at the slit tip, a bright and localized PL spot appears. Scale bar: 1 μm. **c** and **d** share the same color bar clearly showing an increased fluorescence at the tip apex between the electrodes. **e-f**, AFM scans of two representative devices after the DEP process. **e**, A single sphere structure is positioned at the cavity tip using DEP parameters: voltage = 2 V, frequency = 2 MHz, duration = 3 s. **f**, With modified DEP parameters (voltage = 2 V, frequency = 0.5 MHz, duration = 3 s), a larger number of particles, possibly including contaminants, accumulate around the cavity tip. Scale bar: 500 nm.

**Strongly coupled spectra in PL**

Split PL spectra indicating strong coupling can be measured directly after the coupled structures are formed as described above. Many experiments rely on observing split scattering spectra as evidence of strong coupling; however, such features can also arise from alternative mechanisms such as Fano resonances[22,24] or inhomogeneous dielectric environments[29], making the interpretation ambiguous. In contrast, split PL spectra, when combined with careful analysis of the uncoupled states, provide more definitive evidence of strong coupling. Typically, we focus a green excitation laser (power: 1 μW) on a position that simultaneously covers both the plasmonic slit cavity and Qdots, ensuring that the Qdots are primarily excited. To identify the correct position, we record a hyperspectral image and locate the brightest spot by integrating over the spectral range of the Qdot emission. Notably, PL signals from the bare plasmonic slit cavity are only detectable under high excitation power (500 μW), and thus are negligible under our low-power conditions.

Once the Qdots are strongly coupled to the slit cavity, polaritons are formed through coherent energy exchange between excitons and cavity plasmons. This interaction manifests spectrally as a characteristic double-peak feature in the PL spectra, as shown in Figure 4a-c, recorded from three different devices prepared in separate experiments under the same conditions.

To interpret the experimental results, we use the Jaynes–Cummings Hamiltonian to model the interaction between a single plasmonic nanocavity mode and a two-level quantum emitter[30]. Lindblad terms are incorporated into the master equation to account for incoherent pumping and all relevant loss and dephasing channels (Supplementary Information S6). Because each experiment begins with a characterization of the slit cavity, the cavity resonance frequency $\omega_{cav}$ and loss rate $\gamma_{cav}$ can be accurately determined and fixed in the model. In contrast, the Qdot emission frequency $\omega_{qd}$ and linewidth $\gamma_{qd}$, are significantly affected by spectral diffusion at room temperature (Figure 2d). These parameters are therefore adjusted slightly to fit the measured spectra (Supplementary Information S7).

Figure 4a presents five PL spectra acquired with different integration times, along with theoretical fits using the quantum model. Variation of the integration time can provide insights into the effects of spectral diffusion. From the fits, an average coupling strength g = 72.4 ± 2.3 meV is extracted, which clearly satisfies the strong coupling criterion: $2g > (\gamma_{cav} + \gamma_{qd})/2$. Meanwhile, when the integration time varies from 33 ms to 1 s, the splitting feature remains clearly visible, with the Rabi splitting strength nearly unchanged.

The same fitting approach is applied to data sets recorded for different structures, Figure 4b and Figure 4c, where hybrid structures were fabricated using identical procedures. The extracted coupling strengths are $101.4 \pm 1.8$ meV and $123.0 \pm 2.3$ meV, respectively. These variations in different coupled structures are likely to arise from different numbers $N$ of Qdots coupling to the same slit cavity mode, with the effective coupling strength expected to scale as $\sqrt{N}$ in Figure 4d, in agreement with the Tavis–Cummings model[31]. This interpretation is supported by the fact that, in the Qdot aqueous solution, some emitters are present in the form of clusters, which can be verified by the AFM scans of spin-coated Qdots on the coverslip in Figure S4 (Supplementary Information S3). Additional results from other devices are presented in Figure S7d (Supplementary Information), where fluctuations in coupling strength can be ascribed to variations in the dipole orientation of individual Qdots. Furthermore, finite-difference time-domain (FDTD) simulations (Figure S6 in Supplementary Information S5) provide an estimate of the single-emitter coupling strength of 72.8 meV, based on the calculated effective mode volume and the effective dipole moment of a Qdot. In our experiments, we did not observe coupling strengths exceeding those reported here. The reason for this is likely the extent of the spatially confined mode at the nanoslit cavity tip, where the available mode volume restricts the number of Qdots that can strongly couple.

A key advantage of our on-chip platform is the ability to deterministically characterize each slit cavity prior to coupling experiments—something not feasible with systems such as bowtie antennas or particle-on-mirror structures. This pre-characterization eliminates uncertainties due to multiple plasmonic mode interference and enables accurate estimation of coupling strength without uncertainties due to possibly large detunings or uncoupled entities.

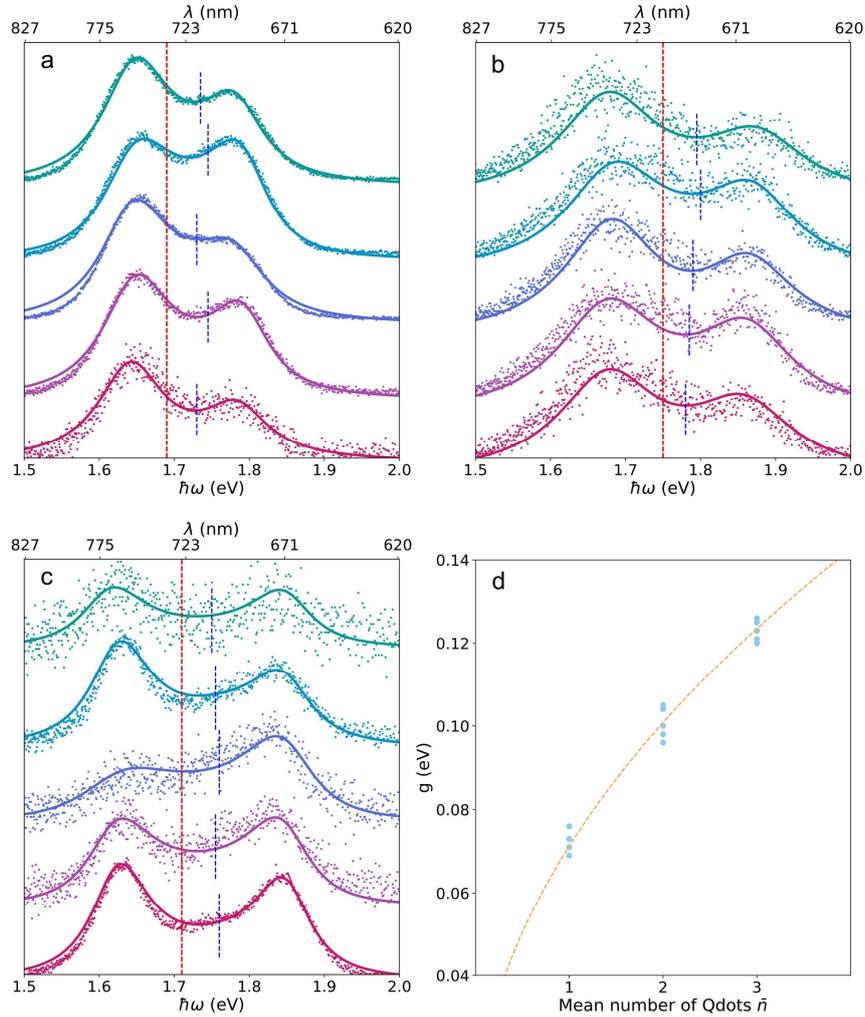

**Figure 4. Split PL spectra recorded from different coupled structures. a-c,** Split PL spectra (dots) recorded from three different coupled structures and the corresponding fit (solid lines) of the quantum model. For each structure, five spectra were recorded with different integration times to verify its reproducibility and to elucidate influences of spectral diffusion. Before fabricating the coupled structures, the resonance of the slit cavity is characterized by PL with high power. The resonance frequency is marked by the red dashed line. The Qdot's resonance is labeled by blue dashed line, which is not fixed in the fitting to accommodate small but finite spectral diffusion at room temperature. The fitting parameters can be found in Table S1, S2, S3, respectively, in Supplementary Information. **d**, The extracted coupling strengths from three different structures are indicated by the dots. The orange line represents the corresponding comparison to theory assuming multiple emitters (indicated by n̄) coupled into a single cavity.

**Electrically connected strongly coupled device**

Benefiting from the electrically connected structure and the gap between the counter electrode and slit tip, we leverage the quantum-confined Stark effect (QCSE) to deliberately tune the resonance of the trapped Qdot by applying a strong external electric field, as illustrated in Figure 5a. QCSE is a well-known phenomenon whereby increasing the electric field reduces the overlap between the electron and hole wavefunctions, resulting in decreased PL intensity and a spectral shift. Typically, linear and quadratic dependence on the electric field is sufficient to describe frequency shifts as[32]:

$$\Delta\omega = -\mu E - \frac{1}{2}\alpha E^2$$

where μ and α are the permanent dipole moment and polarizability of the Qdot. E is the electric field acting on the Qdot, consisting of two contributions: $E_{int}$, the internal electric field arising from the Qdot's charged state (randomly orientated, e.g., through the addition or removal of a surface electron) and $E_{applied}$, the externally applied electric field.

In Figure 5b, we present PL spectra characteristic for strong coupling of a single Qdot to a plasmonic nanoslit cavity (Figure 5a) recorded under different applied DC voltages. When no voltage is applied, the extracted coupling strength is the same as that shown in Figure 4a for an average emitter number $\bar{n} = 1$. As the applied field strength increases, a clear decrease in PL intensity is observed. However, a significant and consistent shift in the Qdot resonance energy is not observed. To further investigate, we analyze multiple split PL spectra under varying applied voltages and extract both the overall PL intensity and the Qdot resonance energy from the spectra, as shown in Figure 5c. While the PL intensity consistently decreases with increasing $E_{applied}$, the resonance shift exhibits no dependence on the electric field, with an overall tuning range of approximately 20 meV—still within the typical range of spectral diffusion. Notably, similar fluctuations in the Qdot resonance are also observed in the absence of any applied voltage, as shown previously in Figure 4a-c. Similar observations are presented in Supplementary Information S8 and Figure S8a–b, where alternating voltages were applied to the hybrid structures. Once again, the resonance shifts show no clear dependence on the applied voltage.

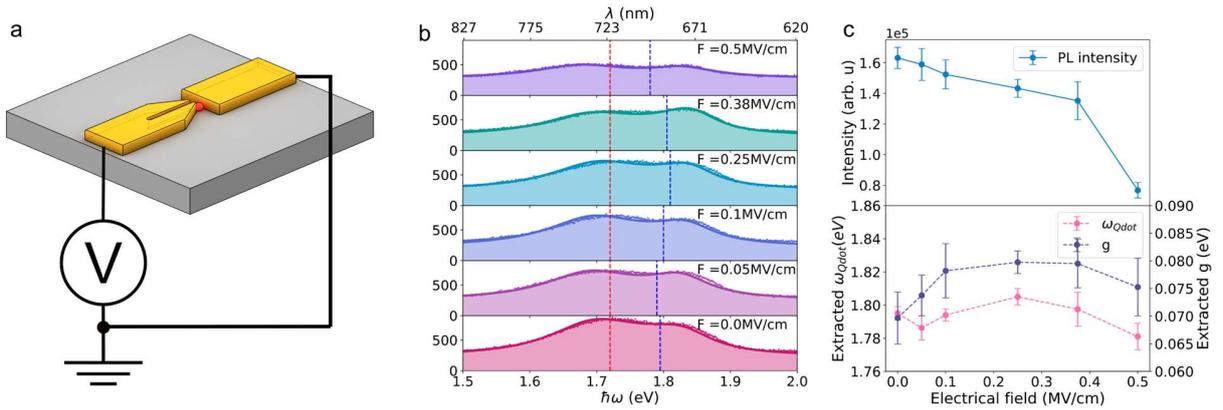

**Figure 5. Electrically connected on-chip strong-coupling devices. a,** Schematic illustration of the electrically connected on-chip strong-coupling device. A DC voltage source is connected to the two electrodes after the coupled structure is formed. **b**, Measured split PL spectra (dots) and corresponding fits from the quantum model (solid lines) under varying applied electric fields from 0 MV/cm to 0.5 MV/cm. The red dashed line indicates the cavity resonance characterized prior to the DEP process, while the blue dashed line marks the Qdot resonance extracted from the fit. **c**, Statistical analysis of PL intensities, extracted Qdot resonances and coupling strength g

across multiple spectra under different applied voltages. PL intensity is calculated by integrating the total counts across the full spectral range.

The Qdot resonance shift induced by the QCSE is typically within 20 meV[18,32–34], consistent with the shifts observed in Figure 5c. However, spectral diffusion driven by fluctuations in the local electric field can span up to 50 meV, making it difficult to clearly resolve the QCSE-induced shift. This limitation could be addressed by using higher quality Qdots with reduced spectral diffusion at room temperature, or by applying higher electric fields to enhance the Stark shift beyond the diffusion range. The latter approach, however, is technically challenging, as the structures become increasingly vulnerable to damage under stronger electric fields.

**Conclusion**

In summary, we demonstrate an on-chip platform for achieving strong light–matter interaction between single or few Qdots and a plasmonic slit cavity. Precise spatial positioning of Qdots at the cavity hotspot is realized using a DEP process integrated with real-time PL feedback. The coupled structures exhibit clear spectral splitting in PL, consistent with predictions of a quantum model assuming coupling of a two-level system with a single mode. By pre-characterizing the plasmonic cavity properties, we reliably extract coupling strengths for a multitude of fabricated devices. Our analysis reveals variations in coupling strength across different hybrid structures, which are likely determined by the number of emitters coupled to the cavity mode. Furthermore, we explore electrical control over the Qdot resonance via the QCSE enabled by the integrated electrodes of the lateral device architecture. While applied electric fields modulate the PL intensity, spectral diffusion at room temperature obscures clear observation of Stark shifts. These results highlight the potential of deterministic, ultra-compact, and electrically tunable quantum plasmonic systems for applications in quantum information processing, tunable photonic devices, and integrable on-chip quantum light source.

**Methods**

**Sample preparation**

Monocrystalline gold flakes are synthesized in solution following established protocols[35–37]. Flakes with suitable thickness are selected based on transmission measurements and transferred onto a substrate pre-patterned with gold electrodes using a standard optical lift-off process (Figure S1a). A gallium focused-ion beam (FIB) is then used to isolate individual electrodes and define a large optical window along with coarse structural features (Figure S1b). High-precision slit cavities and tip gaps are subsequently fabricated using helium ion milling (Orion nanoFab, Zeiss).

Commercial colloidal semiconductor quantum dots (CdSeTe/ZnS Qdot 705 ITK Carboxyl, Q21361MP, Thermo Fisher Scientific) are used as quantum emitters in our experiments. The Qdot solution is diluted in Milli-Q water to achieve appropriate concentrations for different measurements. The aqueous solution was carefully prepared to ensure a uniform distribution of quantum dots across the glass coverslip, avoiding any interaction between different quantum dots. For basic characterization, a 1:6000 dilution is spin-coated onto cleaned microscope coverslips (Gerhard Menzel GmbH). For DEP positioning, a more dilute solution (1:10000) is used.

**Optical setup**

The PL measurement setup is sketched in Figure S2. A 532 nm continuous-wave laser (AIST-NT ROU006) serves as the excitation source and is focused onto the sample using a high-numerical-aperture objective (Nikon CFI P-Apo 100x, NA 1.45). For characterizing single Qdot emission and conducting coupling experiments, an excitation power of 1 µW is typically used, while a higher power of 500 µW is applied to probe gold photoluminescence. The emitted PL signal passes through a dichroic mirror and is routed via a flip mirror either into a spectrometer (HORIBA iHR320) equipped with an EMCCD detector (Andor Newton 970p), or toward a time-resolved photon statistics setup. Independent piezoelectric stages control the sample mount and objective, enabling precise alignment of the laser focus with the sample. Time-correlated single-photon counting (TCSPC) measurements are performed by splitting the PL signal with a 50:50 beam splitter and detecting photons using two avalanche photodiodes (APDs, SPCM-AQR). Photon arrival times are recorded using a field-programmable gate array (FPGA, qutools quTAU H+).

ASSOCIATED CONTENT

The Supporting Information is available.

> FDTD simulations of plasmonic slit cavities, gold photoluminescence fitting procedures, characterization of quantum dots, detailed steps of the DEP process, classical evaluation of coupling strength, quantum modeling and spectral fitting procedures, and additional experiments on electrically connected strongly coupled devices.

AUTHOR INFORMATION

Corresponding Author

**Jin Qin -** Nano-Optics and Biophotonics Group, Experimentelle Physik 5, Physikalisches Institut, Universität Würzburg & and Röntgen Research Center for Complex Material Research, Physics Institute, Am Hubland, Würzburg, D-97074, Germany; Email: jin.qin@uni-wuerzburg.de

**Ortwin Hess** -  School of Physics and CRANN Institute, Trinity College Dublin, Dublin, Ireland

Author Contributions; Email: Ortwin.Hess@tcd.ie

**Bert Hecht -** Nano-Optics and Biophotonics Group, Experimentelle Physik 5, Physikalisches Institut,Universität Würzburg & and Röntgen Research Center for Complex Material Research, Physics Institute, Am Hubland, Würzburg, D-97074, Germany; Email:  hecht@physik.uni-wuerzburg.de

Author

**Benedikt Schurr -** Nano-Optics and Biophotonics Group, Experimentelle Physik 5, Physikalisches Institut,Universität Würzburg & and Röntgen Research Center for Complex Material Research, Physics Institute, Am Hubland, Würzburg, D-97074, Germany

**Patrick Pertsch -** Nano-Optics and Biophotonics Group, Experimentelle Physik 5, Physikalisches Institut,Universität Würzburg & and Röntgen Research Center for Complex Material Research, Physics Institute, Am Hubland, Würzburg, D-97074, Germany

**Daniel Friedrich -** Nano-Optics and Biophotonics Group, Experimentelle Physik 5, Physikalisches Institut,Universität Würzburg & and Röntgen Research Center for Complex Material Research, Physics Institute, Am Hubland, Würzburg, D-97074, Germany

**Max Knopf** - Nano-Optics and Biophotonics Group, Experimentelle Physik 5, Physikalisches Institut,Universität Würzburg & and Röntgen Research Center for Complex Material Research, Physics Institute, Am Hubland, Würzburg, D-97074, Germany

**Saeid Asgarnezhad-Zorgabad** - School of Physics and CRANN Institute, Trinity College Dublin, Dublin, Ireland

**Lars Meschede** - School of Physics and CRANN Institute, Trinity College Dublin, Dublin, Ireland

**Daniel D.A. Clarke** - School of Physics and CRANN Institute, Trinity College Dublin, Dublin, Ireland

**Monika Emmerling** - Nano-Optics and Biophotonics Group, Experimentelle Physik 5, Physikalisches Institut,Universität Würzburg & and Röntgen Research Center for Complex Material Research, Physics Institute, Am Hubland, Würzburg, D-97074, Germany

**Artur Podhorodecki** - Department of Experimental Physics, Wroclaw University of Science and Technology, Wybrzeze Wyspianskiego, Wroclaw, 50-370, Poland.

J.Q. and B.S. contributes equally to this work. B.H. supervised and initiated the experiments. B.S., P.P, M.K. and D.F. fabricated the sample and did the optical characterization. J.Q. processed the data and implemented simulations. J.Q. wrote the manuscript. All authors contributed to discussing the data and commenting on the paper.

Data Availability

The data generated and analyzed in the current study are available from the corresponding author on reasonable request.

Notes

The authors declare no competing financial interest.

ACKNOWLEDGMENT

This project was partly funded within the QuantERA II Programme that has received funding from the European Union's Horizon 2020 research and innovation programme under Grant Agreement

No 101017733, and with DFG (HE5618/12-1), Research Ireland (22/QERA/3821), and NCN (Poland, 2021/03/Y/ST5/00174). We gratefully acknowledge funding by the Deutsche Forschungsgemeinschaft (DFG, German Research Foundation) under Germany's Excellence Strategy through the Würzburg-Dresden Cluster of Excellence on Complexity and Topology in Quantum Matter, ct.qmat (EXC 2147, Project ID ST0462019) as well as through a DFG project (INST 93/959-1 FUGG), a regular project (HE5618/10-1), and a Reinhard-Koselleck project (HE5618/6-1). We further acknowledge funding by the "Bayerische Staatsministerium für Wissenschaft und Kunst", via the program "Grundlagen-orientierte Leuchtturmprojekte für Forschung, Entwicklung und Anwendungen im Bereich Quantenwissenschaften und Quantentechnologien" within the "Munich Quantum Valley" (IQ-Sense).

# Supplementary Information for On-chip plasmonic slit-cavity platform for room-temperature strong coupling with deterministically positioned colloidal quantum dots

Jin Qin,*†‡ Benedikt Schurr,†‡ Patrick Pertsch,† Daniel Friedrich,† Max Knopf,† Saeid Asgarnezhad-Zorgabad,¶ Lars Meschede, ¶ Daniel D.A. Clarke,¶ Monika Emmerling,† Artur Podhorodecki,§ Ortwin Hess, *¶ and Bert Hecht*†

†Nano-Optics and Biophotonics Group, Experimentelle Physik 5, Physikalisches Institut, Universität Würzburg & and Röntgen Research Center for Complex Material Research, Physics Institute, Am Hubland, Würzburg, D-97074, Germany.

¶School of Physics and CRANN Institute, Trinity College Dublin, Dublin, Ireland.

§Department of Experimental Physics, Wroclaw University of Science and Technology, Wybrzeze Wyspianskiego, Wroclaw, 50-370, Poland.

‡ These two authors contribute equally.

* jin.qin@uni-wuerzburg.de; Ortwin.Hess@tcd.ie; hecht@physik.uni-wuerzburg.de

**S1 Simulations**

The quasinormal mode (QNM) analysis of the plasmonic slit cavity was performed using the COMSOL mode solver. The dielectric function of monocrystalline gold was taken from Olmon et al.[1], and a refractive index of 1.52 was used for the glass substrate. Perfectly matched layers were applied to the simulation boundaries to absorb outgoing electromagnetic waves and eliminate artificial reflections. Numerical artifacts were removed to ensure accurate results. Figure 1c was generated by summing the contributions of the second- and third-order QNMs, modeled using Lorentz oscillators.

**S2 Gold PL fitting model**

The spectrum of the plasmonic slit cavity is characterized by analyzing the intrinsic linear photoluminescence (PL) of gold, excited using a 532 nm continuous-wave laser diode at an intensity of $4.4 \times 10^9 \text{W/m}^2$[2]. Owing to local field enhancement and the increased local density of optical states (LDOS) associated with the plasmonic mode, the gold PL background develops a distinct peak, which becomes clearly visible at sufficiently high excitation power. The resonance energy and quality factor Q of the slit cavity resonance are extracted from the gold PL spectrum using a composite fitting function. This model includes an exponential decay to account for the broad unstructured background and a Lorentzian to describe the cavity mode:

$$f(\omega) = \frac{A_a e^{-b\omega}}{\gamma_a[1+4(\omega-\omega_a)^2/\gamma_a^2]} + A_a e^{-b\omega} + c \tag{S1}$$

Here, $A_a$, $\gamma_a$, and $\omega_a$ represent the amplitude, full width at half maximum (FWHM), and resonance frequency of the slit cavity, respectively. The extracted quality factor is 15.8, which closely matches the value obtained from simulations 20.

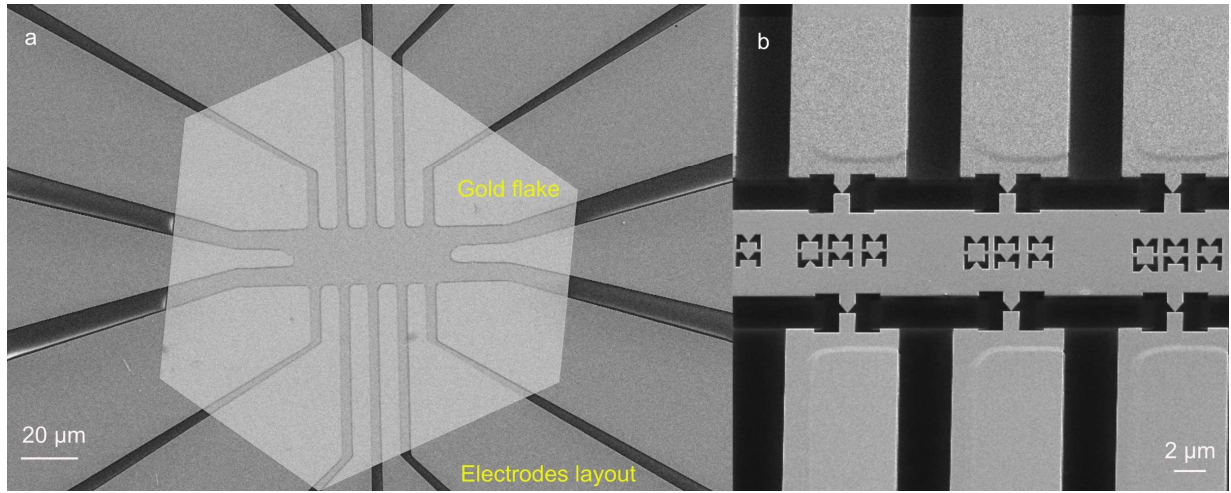

**Figure S1 SEM images of the fabrication process for the on-chip plasmonic device. a**, A monocrystalline gold flake is transferred onto the pre-patterned electrode layout. **b**, Gallium FIB processing is used to electrically isolate the electrode fingers and define the large optical window.

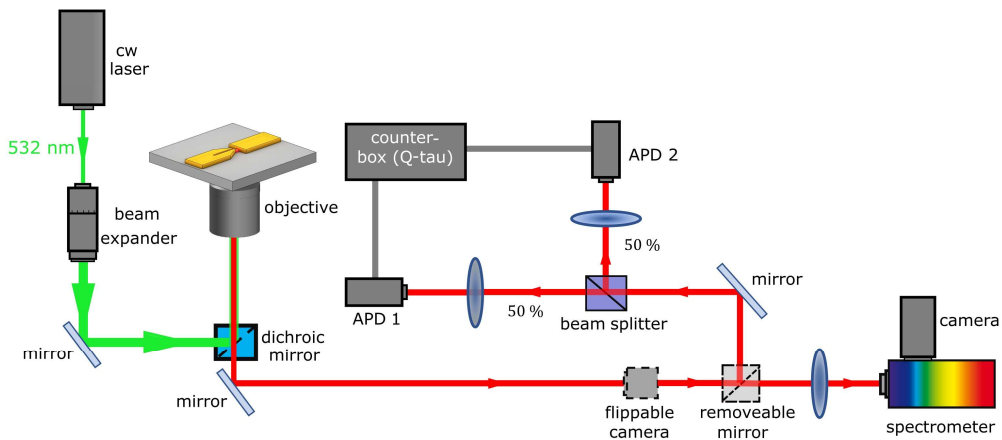

**Figure S2 Schematic of PL measurement setup.**

**S3 Quantum Dot**

To verify the presence of clusters in the Qdot aqueous solution, we spin-coated the solution onto a coverslip and performed AFM scans, as shown in Figure S4. Only a small fraction of the features corresponds to individual Qdots, which exhibit small dimensions comparable to the one in Figure 2a. In contrast, some features display larger dimensions, indicating that they are likely composed of multiple Qdots.

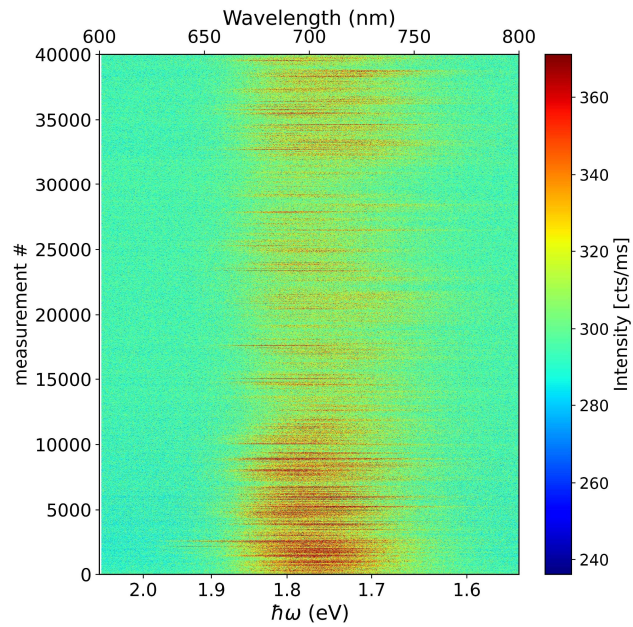

**Figure S3 Time traced PL spectra from a single Qdot.** Each spectrum is taken within an integration time of 33 ms.

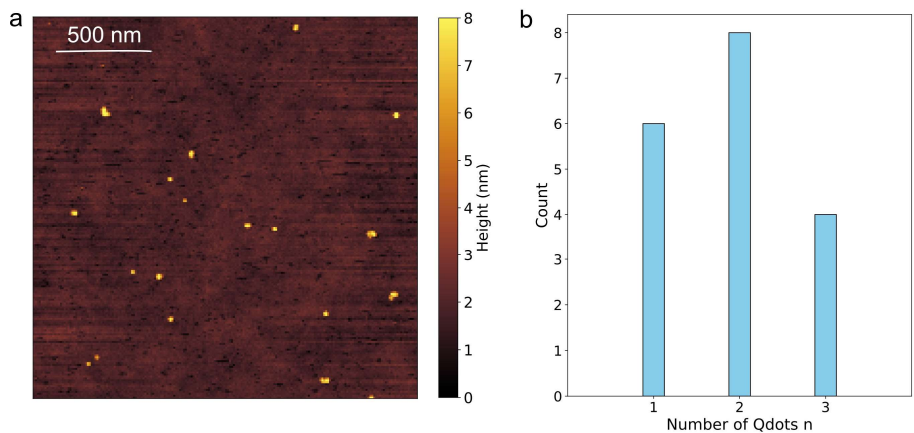

**Figure S4 AFM scans of Qdots spin-coated on a coverslip. a**, AFM image over an area of 2 μm by 2 μm. **b**, Statistics of the number of Qdots per cluster.

## S4 Dielectrophoresis process

Dielectrophoresis (DEP) occurs when a polarizable particle is placed in a spatially non-uniform electric field, resulting in a net force due to the field gradient. The time-averaged DEP force acting on a spherical particle can be expressed as[3]:

$$\langle F_{DEP}\rangle = 2\pi\epsilon_1 R^3 \left[\frac{\epsilon_2-\epsilon_1}{\epsilon_2+2\epsilon_1} + \frac{3(\epsilon_1\sigma_2-\epsilon_2\sigma_1)}{\tau_m(\sigma_2+2\sigma_1)^2(1+\omega^2\tau_m^2)^2}\right]\nabla E_{rms}^2 \qquad (S2)$$

where $\epsilon_1$ ($\epsilon_2$) and $\sigma_1$ ($\sigma_2$) denote the permittivity and conductivity of the surrounding medium (particle), respectively. R is the particle radius, $\omega$ is the frequency of the applied field, and $E_{rms}$ is the root-mean-square magnitude of the electric field. The term $\tau_m = \frac{\epsilon_2+\epsilon_1}{\sigma_2+2\sigma_1}$ is the Maxwell-Wagner relaxation time, characterizing the decay of the dipolar charge distribution at the particle interface.

Equation S2 indicates that the magnitude of the DEP force can be tuned by adjusting the frequency $\omega$, providing a route to optimize the trapping process. If $\langle F_{DEP}\rangle$ is too large, many particles are pulled toward the target, promoting clustering. Increasing $\omega$ reduces the force, allowing more controlled placement of particles—especially when combined with real-time feedback.

The detailed DEP procedure is illustrated in Figure S5. First, the Qdot solution is ultrasonicated for one minute to disperse aggregates, then diluted in Milli-Q water at a 1:10000 ratio, followed by another one-minute ultrasonication. A 5 μL droplet is deposited on the structured gold flake, fully covering the active area. Two micromanipulators contact the electrodes, which are connected to a function generator (DS345, Stanford Research Systems). A 532 nm green laser is focused onto the slit cavity, and the PL signal is monitored in real-time using a sensitive avalanche photodiode (APD).

Upon applying an AC electric field, Qdots are drawn to the cavity gap. A sudden increase in photon counts indicates successful capture, triggering immediate shutdown of the function generator to prevent additional accumulation. The laser spot and micromanipulators are then moved to the next cavity, and the process is repeated. To avoid drying-up, water is added from time to time. After the DEP process is complete, the droplet is rinsed off with Milli-Q water followed by ethanol, and the sample is then thoroughly dried.

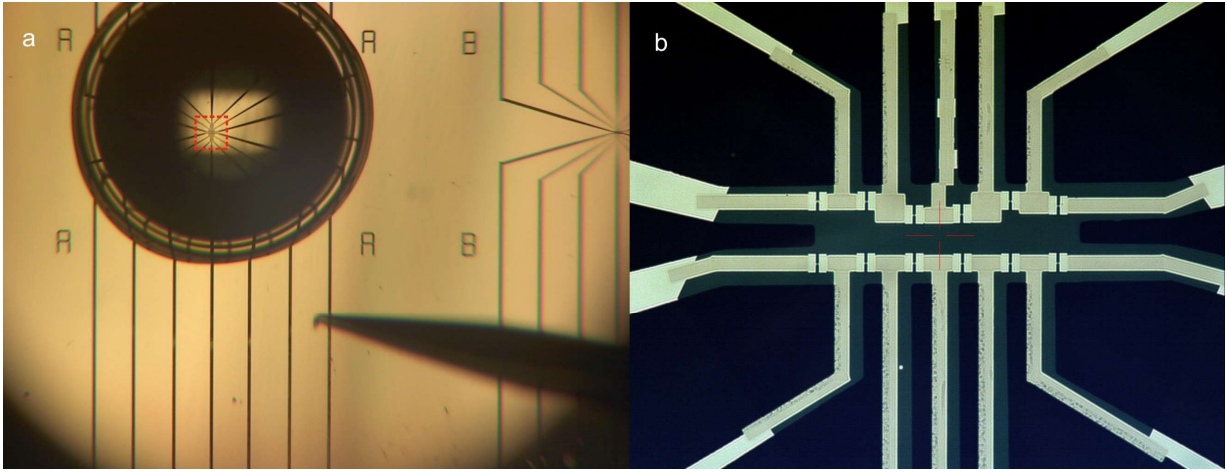

**Figure S5 Microscopic images of the DEP process. a**, A 5 µL aqueous droplet containing Qdots is placed on top of the structures. Micromanipulators are used to contact the two electrodes from above, while PL spectra are monitored from below using a high-NA objective. **b**, A zoomed-in view showing the transmission image of the structured gold flake under the droplet (corresponding to the red dashed box in **a**).

**S5 Classical evaluation**

FDTD simulations were performed to investigate the coupled states between a plasmonic slit cavity and a single quantum emitter. The dielectric function of single-crystalline gold was taken from Olmon et al.[1]. To model the quantum dot exciton state, its permittivity was described by a Lorentzian function,

$$\varepsilon_{Qdot}(\omega) = \varepsilon_\infty + f\omega_0^2/(\omega_0^2 - \omega^2 - i\gamma_0\omega) \tag{S3}$$

following the Refence[4]. Here, $\varepsilon_\infty$ is the high-frequency contribution of the CdSe/ZnS quantum dot matrix dielectric function, with a value of 5. The oscillator strength was set to 0.3, the lowest excitonic transition was placed at 1.76 eV, and the exciton linewidth was 140 meV.

The coupling energy g between the plasmonic slit cavity and the quantum dot is given by the scalar product of the quantum dot dipole moment $\mu$ and the vacuum field amplitude $E_0$ at position $\boldsymbol{r}$,

$$E_0(\boldsymbol{r}) = \sqrt{\frac{\hbar\omega}{2\varepsilon_0 V_{eff}(\boldsymbol{r})}} \tag{S4}$$

where $\mu$ is the dipole moment of the quantum dot, $\hbar\omega$ is the photon energy, $\varepsilon_0$ is the permittivity of free space, and $V_{eff}(\boldsymbol{r})$ is the effective mode volume of the slit cavity. The effective mode volume was calculated using quasi-normal modes (QNMs) with complex frequencies following Sauvan et al.[2,5]. Based on these simulations, the single-emitter coupling strength is estimated to be 72.8 meV.

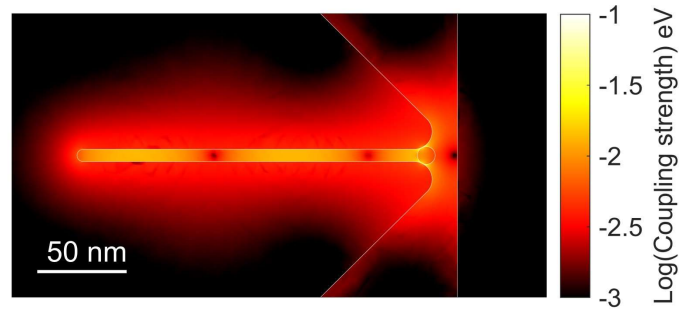

**Figure S6 Map of the coupling strength g in eV (log scale) obtained from FDTD simulations.** The quantum dot is assigned with a dipole moment of 9 Debye. To obtain the spatially dependent coupling strength, the effective mode volume was multiplied by a dimensionless overlap factor $\eta = \max\left(\left|\vec{E}\right|^2\right)/\left|\vec{E}\right|^2$. Scale bar: 50 nm.

**S6 Quantum model**

Light–matter strong coupling can be described by the Jaynes–Cummings Hamiltonian (with $\hbar = 1$)[6]:

$$H = \omega_{cav} a^\dagger a + \omega_{qd} \sigma^\dagger \sigma + g(a^\dagger \sigma + \sigma^\dagger a) \qquad \text{S5}$$

where $a$ ($a^\dagger$) are the annihilation (creation) operators of a single bosonic cavity mode with energy $\omega_{cav}$, and $\sigma^\dagger(\sigma)$ are the raising (lowering) operators of a two-level system with transition energy $\omega_{qd}$. The two systems—the optical cavity and Qdot—are coupled with strength g.

To analyze the dynamics of this hybrid system, we solve the master equation for the density matrix $\rho$:

$$\frac{d}{dt}\rho = -i[H,\rho] + \sum_i \mathcal{L}_i(\rho) \qquad \text{S6}$$

where $\mathcal{L}_i$ are Lindblad superoperators representing various dissipative processes, including radiative decay ($\gamma_{cav}, \gamma_{qd}$), incoherent pumping ($P_{cav}, P_{qd}$), and pure dephasing ($\gamma_\phi$). These are expressed as:

$$\sum_i \mathcal{L}_i(\rho) = \frac{\gamma_{cav}}{2}(2a\rho a^\dagger - \{a^\dagger a, \rho\}) + \frac{\gamma_{qd}}{2}(2\sigma\rho\sigma^\dagger - \{\sigma^\dagger\sigma, \rho\}) +$$

$$\frac{P_{cav}}{2}(2a^\dagger\rho a - \{aa^\dagger, \rho\}) + \frac{P_{qd}}{2}(2\sigma^\dagger\rho\sigma - \{\sigma\sigma^\dagger, \rho\}) +$$

$$\frac{\gamma_\phi}{2}(\sigma_z\rho\sigma_z - \rho). \qquad \text{S7}$$

We assume that cavity emission dominates in the hybrid system, allowing us to neglect direct emission from the quantum dot[7]. Under this assumption, the emission spectrum $S(\omega)$ is calculated as:

$$S(\omega) \propto \Re(\int_0^\infty \langle a^\dagger(\tau)a(0)\rangle e^{-i\omega}\, d\tau) \qquad \text{S8}$$

Numerical simulations are performed using the QuTiP (Quantum Toolbox in Python) package[8].

## S7 Spectral fitting procedure

Since the bare slit cavity is characterized prior to the DEP process, the cavity resonance frequency $\omega_{cav}$ and cavity loss rate $\gamma_{cav}$ can be extracted by fitting the intrinsic gold PL spectrum, as shown in Figure 1c. These parameters are kept fixed during the fitting of the coupled spectra. The Qdot resonance frequency $\omega_{qd}$, however, varies due to spectral diffusion at room temperature and is adjusted accordingly during the fitting procedure.

The total linewidth of the Qdot emission consists of two contributions: the radiative decay rate $\gamma_{qd}$ and the pure dephasing rate $\gamma_\phi$. From time-resolved photon statistics measurements, the Qdot lifetime $\tau$ is determined to be 33 ns, which corresponds to a radiative decay rate of $\gamma_{qd} = 20\,neV$, based on the relation $\gamma_\sigma \tau = \hbar$. The dominant contribution to the broad linewidth of the Qdot arises from pure dephasing ($\gamma_\phi$), which fluctuates over time as shown in Figure 2d.

The detailed fitting parameters used in the simulations and spectral fitting are summarized in Tables S1, S2, and S3.

**Table S1 Fitting parameters for Figure 4a with quantum model.**

| $\omega_{cav}$(eV) | g(meV) | $\omega_{qd}$(eV) | $\gamma_{cav}$(meV) | $P_{cav}$(meV) | $\gamma_{qd}$(neV) | $P_{qd}$(meV) | $\gamma_\phi$(meV) |
|---|---|---|---|---|---|---|---|
| 1.69 | 71.0 | 1.73 | 128.9 | 17.8 | 20.0 | 0.16 | 55.7 |
| 1.69 | 73.0 | 1.745 | 128.9 | 10.1 | 20.0 | 5.0 | 55.7 |
| 1.69 | 73.0 | 1.73 | 128.9 | 20.7 | 20.0 | 4.4 | 55.7 |
| 1.69 | 76.0 | 1.745 | 128.9 | 5.1 | 20.0 | 8.5 | 64.0 |
| 1.69 | 69.0 | 1.735 | 128.9 | 12.8 | 20.0 | 3.9 | 50.9 |

**Table S2 Fitting parameters for Figure 4b with quantum model.**

| $\omega_{cav}$(eV) | g(meV) | $\omega_{qd}$(eV) | $\gamma_{cav}$(meV) | $P_{cav}$(meV) | $\gamma_{qd}$(neV) | $P_{qd}$(meV) | $\gamma_\phi$(meV) |
|---|---|---|---|---|---|---|---|
| 1.75 | 98.0 | 1.78 | 206.9 | 38.8 | 20.0 | 0 | 71.0 |
| 1.75 | 100.0 | 1.785 | 206.9 | 23.9 | 20.0 | 4.2 | 77.0 |
| 1.75 | 96.0 | 1.79 | 206.9 | 37.2 | 20.0 | 0 | 55.7 |
| 1.75 | 104.0 | 1.8 | 206.9 | 27.7 | 20.0 | 9.3 | 55.7 |
| 1.75 | 105.0 | 1.795 | 206.9 | 38.7 | 20.0 | 0 | 66.5 |

**Table S3 Fitting parameters for Figure 4c with quantum model.**

| $\omega_{cav}$(eV) | g(meV) | $\omega_{qd}$(eV) | $\gamma_{cav}$(meV) | $P_{cav}$(meV) | $\gamma_{qd}$(neV) | $P_{qd}$(meV) | $\gamma_{\phi}$(meV) |
|---|---|---|---|---|---|---|---|
| **1.71** | 121.0 | 1.76 | 109.8 | 11.1 | 20.0 | 11.2 | 23.9 |
| **1.71** | 123.0 | 1.755 | 109.8 | 3.9 | 20.0 | 18.0 | 49.2 |
| **1.71** | 126.0 | 1.76 | 109.8 | 0.1 | 20.0 | 8.5 | 32.9 |
| **1.71** | 120.0 | 1.755 | 109.8 | 17.0 | 20.0 | 7.4 | 28.2 |
| **1.71** | 125.0 | 1.75 | 109.8 | 4.9 | 20.0 | 10.9 | 55.7 |

In Figure S7a–c, we present additional AFM scans of coupled devices fabricated using the same DEP process with feedback control. Typically, AFM characterization is performed after spectral splitting is observed in the PL measurements, serving as further confirmation of the coupled behavior. Furthermore, Figure S7d shows coupling strengths extracted from PL spectra of ten different coupled structures. While some fluctuations are observed across devices, these can be explained by variations in the dipole orientation of the Qdots and by differences in the number of Qdots coupled to the plasmonic slit cavity mode.

**S8 Another dataset for an electrically connected strongly coupled device**

Another demonstration that spectral diffusion plays a dominant role in QCSE experiments at room temperature is provided in Figure S8a-b, where we perform a voltage sweep sequence: starting from zero, increasing to a positive field, returning to zero, and then applying a negative field, as illustrated in Figure S8a. The corresponding PL intensities and extracted Qdot resonances are shown in Figure S8b. A significant drop in PL intensity is observed under a positive applied electric field, which is partially recovered upon returning to zero voltage. Conversely, under a negative applied field, the PL intensity remains relatively stable. This asymmetric response can likely be attributed to the presence of a local internal electric field $E_{int}$, as schematically depicted in Figure S8c. When $E_{applied}$ and $E_{int}$ are in the same direction, the total field increases, further reducing electron-hole wavefunction overlaps and decreasing PL intensity. In contrast, when the two fields are oppositely directed, the effective field is reduced or even canceled, potentially neutralizing the Qdot and enhancing PL emission. Particularly, the PL intensities at zero applied voltage (measurements #0, #2, #4, #6, and #8) exhibit significant fluctuations, indicating that the internal electric field $E_{int}$ varies randomly over time and can be comparable in magnitude to the externally applied field $E_{int}$[9,10]. Additionally, the resonance tunability remains difficult to quantify, as the spectral fluctuations observed at zero applied field are larger than those induced by the applied voltage, further obscuring clear Stark shifts.

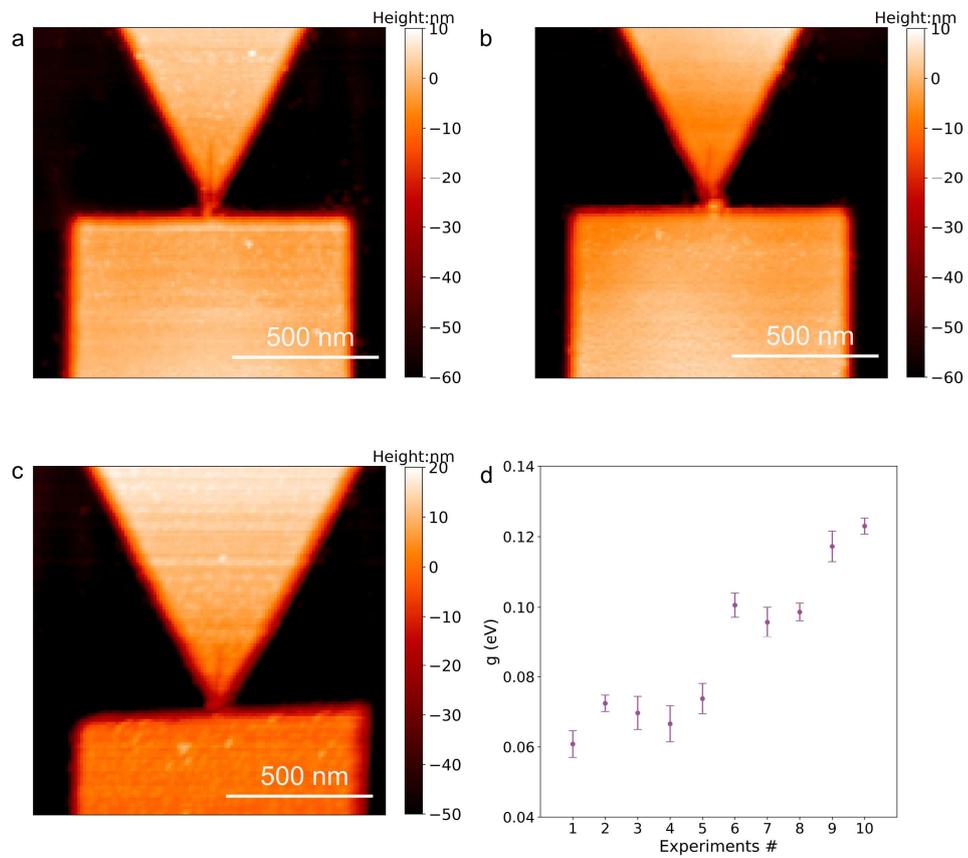

**Figure S7 AFM Characterization and Coupling Strength Statistics of Coupled Devices**. **a–c**, AFM scans of three additional coupled devices. **d**, Coupling strengths extracted from ten other strongly coupled devices.

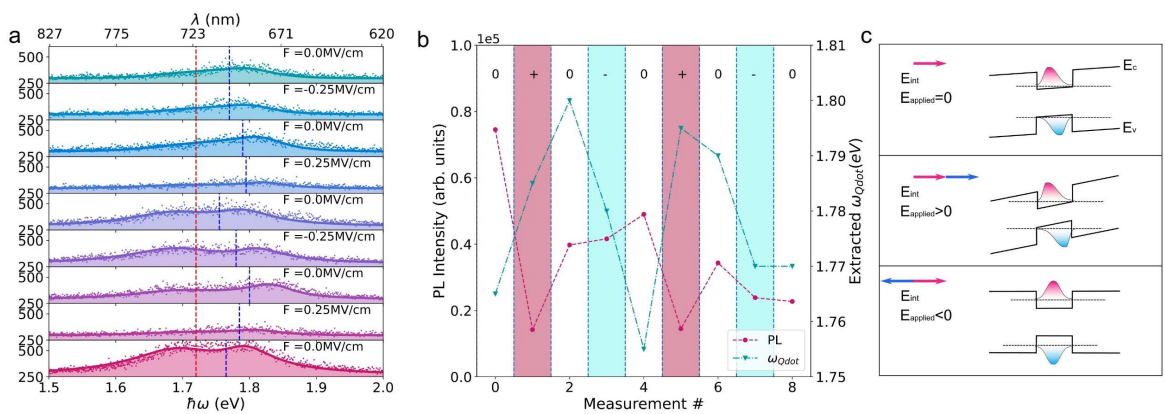

**Figure S8 Additional split PL spectra with different applied voltages. a**, Recorded split PL spectra using a voltage sequence of zero, positive, zero, and negative electric fields. **b**, Corresponding PL intensities and extracted Qdot resonances as a function of the applied electric field. **c**, Band diagrams of the Qdot showing electron and hole wavefunction overlap under the influence of the $E_{applied}$ and $E_{int}$. In the top panel, $E_{applied} = 0$, and the fluctuating $E_{int}$ causes spectral diffusion with varied PL intensities. In the middle and bottom panels, the direction of $E_{applied}$ is reversed, leading to different effective fields and thus varying PL intensities due to changes in wavefunction overlap.